\documentclass[reprint,aps,prx,amsmath,amssymb,superscriptaddress,floatfix,citeautoscript]{revtex4-2}
\usepackage{amsmath}
\usepackage{graphicx}
\usepackage{color}
\usepackage{multirow}
\usepackage[caption = false]{subfig}
\usepackage{array}
\usepackage{threeparttable}
\usepackage{mathtools}
\usepackage{siunitx}
\usepackage{centernot}
\usepackage{tikz}
\DeclareUnicodeCharacter{2212}{-}
\usepackage[normalem]{ulem}
\useunder{\uline}{\ul}{}

\usepackage[colorlinks=true,allcolors=blue]{hyperref}

\usepackage{braket}

\usepackage{txfonts}

\begin{document}
\title{Toward linear scaling auxiliary field quantum Monte Carlo with local natural orbitals}

\author{Jo S. Kurian}
%\email{jokurian12@gmail.com}
\affiliation{Department of Chemistry, University of Colorado, Boulder, CO 80302, USA}

\author{Hong-Zhou Ye}
\affiliation{Department of Chemistry, Columbia University, New York, NY 10027, USA}

\author{Ankit Mahajan}
\affiliation{Department of Chemistry, University of Colorado, Boulder, CO 80302, USA}
\affiliation{Department of Chemistry, Columbia University, New York, NY 10027, USA}

\author{Timothy C. Berkelbach}
\email{tim.berkelbach@gmail.com}
\affiliation{Department of Chemistry, Columbia University, New York, NY 10027, USA}

\author{Sandeep Sharma}
\email{sanshar@gmail.com}
\affiliation{Department of Chemistry, University of Colorado, Boulder, CO 80302, USA}

\begin{abstract}
We develop a local correlation variant of auxiliary field quantum Monte Carlo (AFQMC) that is based on local natural orbitals (LNO-AFQMC). 
In LNO-AFQMC, independent AFQMC calculations are performed for each localized occupied orbital using a truncated set of tailored orbitals.
Because the size of this space does not grow with system size for a target accuracy, the method has linear scaling.
Applying LNO-AFQMC to molecular problems containing a few hundred to a thousand orbitals, we demonstrate convergence of total energies with significantly reduced costs.
The savings are more significant for larger systems and larger basis sets. 
However, even for our smallest system studied, we find that LNO-AFQMC is cheaper than canonical AFQMC, in contrast with many other reduced-scaling methods. Perhaps most significantly, we show that energy differences converge much more quickly than total energies, making the method ideal for applications in chemistry and material science.
Our work paves the way for linear scaling AFQMC calculations of strongly correlated systems, which would have a transformative effect on ab initio quantum chemistry.
\end{abstract}
\maketitle

\section{Introduction}
The primary objective of the field of electronic structure theory is to develop cost effective and accurate methods that can be applied to challenging problems with ease. 
Existing methods differ in the approximations they use to solve the Schr\"{o}dinger equation, and often the choice of method is limited by available computational resources. 
Although Kohn-Sham density functional theory (DFT) method is the most widely used method, its accuracy is difficult to improve systematically~\cite{Yang2012}. 
The coupled cluster (CC) method with single, double, and perturbative triple excitations [CCSD(T)]~\cite{Bartlett07RMP} is widely regarded as the ``gold standard'' for its accuracy when applied to main group chemistry. 
However, the computational cost of canonical CCSD(T) scales with system size $N$ as $N^7$, which limits its practical application to small systems. 
Moreover, CCSD(T) fails catastrophically for strongly correlated systems.\\ 
The phaseless auxiliary field quantum Monte Carlo (AFQMC)~\cite{marioZhangAbInitioAFQMC} method has gained popularity due to its comparatively low $N^4$ computational scaling (i.e., the same as mean-field theory) and impressive accuracy even for strongly correlated systems~\cite{hubbardBenchmark2015,hydrogenBenchmark2017,transitionMetalOxides2020,Shee2020afqmc,Anderson2020afqmc,Friesner2022afqmc_benchmark,Lee2020FullereneFePorphyrin,sukurma2023benchmark, Sharma2022afqmc}. 
AFQMC is a descendent of the determinantal QMC method~\cite{Blankenbeckler1981,DQMCReview2017} and was extensively developed for electronic structure by Zhang and co-workers~\cite{Zhang2006afqmc}. 
Recent algorithmic developments aimed at reducing the cost of AFQMC include the use of low-rank Coulomb integrals~\cite{Morales2019afqmc,Chan2019afqmc,Friesner2022afqmc}, stochastic resolution of identity~\cite{Reichman2020afqmc}, and local trial states~\cite{pham2023scalable}.

Even though AFQMC has a better formal scaling than traditional correlated methods widely used in quantum chemistry, it can still be computationally infeasible for larger systems due to a large prefactor. 
In this work, we overcome this limitation of AFQMC through the use of local correlation. 
The idea to use locality to reduce the cost of correlated calculations dates back to the early seventies~\cite{Meyer1971}, but local correlation was developed in its modern form by Pulay and Saeb\o\ in the eighties~\cite{PULAY1983151,Pulay1986OrbitalinvariantFA,SAEBO198513,Pulay1987JChPh..86..914S}.
In this pioneering work, it was recognized that the electron correlation decays rapidly with the distance between localized orbitals.
This observation underpins so-called ``direct'' local correlation methods, in which a single calculation (almost exclusively perturbation theory or coupled cluster theory) is performed in a basis of localized orbitals, permitting the discarding of small terms and resulting in a calculation with linear asymptotic scaling.
Methods of this type have been extensively developed by many researchers~\cite{WernerSchutz1999,Scuseria1999,WernerSchutz2000,Schutz2000CC,Schutz2002,HeadGordon2005,Schutz2006DF,ShutzDF2007}.
The revival of pair natural orbital based local correlation methods~\cite{Edmiston65,AhlrichsPNO,MeyerPNO73} by Neese, Valeev, and co-workers~\cite{Neese2013,Neese2013-3exPNO,Neese2015,NeesePNO2016} was a major advance of the last decade,
and many related variants have since been developed~\cite{Werner2015PNO,Dornbach2015PNO,Werner2015PNOLNMP2,Hattig2016PNO,Hattig2012PNOMP2,Hattig2013OSVPNO,Hattig2014PNOMMP2,Chan2011LMP2,ChanL2012LMP2,ChanWerner2012LCC,WernerChan2013OSVPNO}. 
Within QMC, a linear scaling version of diffusion Monte Carlo was developed that used Wannier functions to calculate wave function overlaps with linear scaling~\cite{Williamson2001}. 
This work was later extended to improve the scaling of local energy evaluation~\cite{Manten2003,Alfe2004,Kussmann2008}.

A different approach to local correlation, often called fragment-based methods, was initiated by F\"{o}rner and co-workers~\cite{FORNER1985251,FORNER198721}. 
These methods partition the problem into subsystems, and a separate calculation is performed on each subsystem. %, subsystem pairs, subsystem trimers etc until convergence. 
Many variants exist, including the incremental method~\cite{Stoll1992,Dolg1997,Paulus2004,Birkenheuer2004,Stoll1998,Stoll2010, janus2021,zimmerman21}, divide and conquer~\cite{Li2004DC,Nakai2008DC}, divide-expand-consolidate~\cite{Jorgensen2015DEC,Jorgensen2010DEC,Kjergaard2017DEC}, and cluster-in-molecule (CIM)~\cite{Li02JCC,Li09JCP}. 
A significant advantage of such fragment methods over direct methods is that they are easy to implement, can be easily modified for different electron correlation methods, and parallelize trivially. 
In this work, we combine AFQMC with the CIM approach, in particular the local natural orbital (LNO) based variant recently proposed by K\'{a}llay and co-workers~\cite{Rolik11JCP,Rolik13JCP,Nagy17JCP,Nagy18JCTC}. 
Importantly, we show that our fragment based approach is cheaper than the canonical one even for small system sizes.

The rest of the paper is organized as follows. 
In Section~\ref{sec:theory}, we present the basic theory of both LNO based local correlation methods and canonical AFQMC with an emphasis on those aspects of AFQMC that will be modified to develop LNO-AFQMC. 
We end this section with an analysis of the computational scaling of LNO-AFQMC.
In Section~\ref{sec:results}, we compare the efficiency of LNO-AFQMC for calculating the absolute energies and relative energies compared to canonical AFQMC for molecules and reactions of different sizes using various basis sets. 
In Section~\ref{sec:conc}, we conclude and suggest future research directions. 
 
\section{Theory}
\label{sec:theory}

In this work, we consider only closed-shell molecules described by a 
spin-restricted Hartree-Fock (HF) reference determinant $\ket{\Phi_0}$, 
with canonical HF orbitals $\psi_p$, orbital energies $\epsilon_p$, and total
energy $E_\mathrm{HF}$.
We use $i,j,k$ for $N_\mathrm{o}$ occupied orbitals, $a,b,c$ for $N_\mathrm{v}$ virtual orbitals, 
and $p,q,r,s$ for $N$ unspecified molecular orbitals.
In this basis, the electronic Hamiltonian is
\begin{equation}
    H
        = \sum_{pq,\sigma}^{N} h_{pq} a_{p\sigma}^{\dagger} a_{q\sigma} +
        \frac{1}{2} \sum_{pqrs,\sigma\sigma'}^{N} V_{prqs} a_{p\sigma}^{\dagger} a_{q\sigma'}^{\dagger} a_{s\sigma'} a_{r\sigma}
        %= H_1 + H_2
\end{equation}
with $V_{pqrs} = (pq|rs)$ in $(11|22)$ notation.

\subsection{LNO coupled cluster theory}

For completeness, we describe the basics of the LNO-CCSD method~\cite{Rolik11JCP,Rolik13JCP,Nagy17JCP,Nagy18JCTC}.
In this approach, the correlation energy 
is obtained by left projection onto the HF determinant,
\begin{equation}    \label{eq:eccsd_def}
    E_{\mathrm{c}}
        = \braket{\Phi_0 | \bar{H}-E_\mathrm{HF} | \Phi_0}
        = \sum_{ijab} T_{iajb} (2V_{iajb} - V_{ibja})
        = \sum_I E_I
\end{equation}
where $\bar{H}$ is the similarity-transformed Hamiltonian,
$T_{iajb} = t_{iajb} + t_{ia}t_{jb}$,
and $t_{ia}, t_{iajb}$ are the CC single and double amplitudes~\cite{Bartlett07RMP}.
In the final equality of Eq.~(\ref{eq:eccsd_def}), we have recognized
that the energy expression is invariant to unitary rotations of occupied
and virtual orbitals and associated an energy contribution to each
rotated occupied orbital
\begin{equation}
\label{eq:occ_rot}
    \phi_{I}
        = \sum_{i} U_{iI} \psi_i.
\end{equation}

In LNO methods, the unitary transformation of occupied orbitals~(\ref{eq:occ_rot}) 
is chosen to spatially localize the orbitals.  
For each $\phi_I$, one constructs a local active space
$\mathcal{P}_I$ by augmenting $\phi_I$ with selected LNOs (both occupied and virtual) from second-order
M{\o}ller-Plesset perturbation theory (MP2).  
Specifically,
one computes the occupied-occupied and the virtual-virtual blocks of the MP2
density matrix
\begin{align}
    D^{I}_{ij}
        % &= \frac{1}{2}\sum_{ab} t^{(1)}_{iaIb} \big[ 2 t^{(1)}_{jaIb} - t^{(1)}_{Iajb} \big] - U_{iI} U_{jI}
        &= \sum_{ab} t^{(1)}_{iaIb} \big[ 2 t^{(1)}_{jaIb} - t^{(1)}_{Iajb} \big]
        \label{eq:DIij}  \\
    D^{I}_{ab}
        &= \sum_{jc} t^{(1)}_{Iajc} \big[ 2 t^{(1)}_{Ibjc} - t^{(1)}_{jbIc} \big]
        \label{eq:DIab}
\end{align}
where
\begin{equation}    \label{eq:lno_approx_MP2t}
    t^{(1)}_{Iajb} = \frac{(Ia|jb)}{\tilde{\epsilon}_I + \epsilon_{j} - \epsilon_{a} - \epsilon_{b}}
\end{equation}
is an approximate MP2 amplitude with $\tilde{\epsilon}_I = \braket{\phi_I| f |\phi_I}$,
and $f$ is the Fock operator.
Diagonalizing the virtual-virtual block
\begin{equation}
    % D^{I}_{ij}
    %     = \sum_{k} \xi_k^{I} X_{ik}^{I} X_{jk}^{I},
    % \quad{}
    D^{I}_{ab}
        = \sum_{c} \xi_c^{I} X_{ac}^{I} X_{bc}^{I},
\end{equation}
gives the virtual LNOs associated with $\phi_I$, i.e., $\phi_b = \sum_c X_{cb}^I\psi_c$.
For the occupied LNOs, we follow Ref.~\onlinecite{Rolik11JCP} and diagonalize
\begin{equation}    \label{eq:lno_occ}
    \tilde{D}^{I}_{ij}
        = \sum_{kl} Q_{ik}^{I} D_{kl}^{I} Q_{lj}^{I}
        = \sum_{k}^{N_{\textrm{o}}-1} \xi_k^{I} X_{ik}^{I} X_{jk}^{I}
\end{equation}
where $Q_{ij}^{I} = \delta_{ij} - U_{iI} U_{jI}$ projects out $\phi_I$ from $D_{ij}^{I}$ to prevent it from mixing with other occupied orbitals,
giving the occupied LNOs $\phi_j = \sum_k X_{kj}^I \psi_k$.
The eigenvalues $\xi^I_p$, which are between 2 and 0, quantify the importance
of a given LNO to the electron correlation of localized orbital $\phi_I$.
In practice, we construct the local active space $\mathcal{P}_I$ by keeping
those LNOs satisfying

\begin{equation}    \label{eq:lno_truncation}
    \xi_i \geq \epsilon_{\textrm{o}},
    \quad{}
    \xi_a \geq \epsilon_{\textrm{v}},   
\end{equation}
for some user-selected thresholds $\epsilon_{\textrm{o}}$ and $\epsilon_{\textrm{v}}$,
producing $n=n_\mathrm{o}+n_\mathrm{v}$ orbitals. Typically, $n$ is much less than $N$ and
does not increase with system size for a targeted level of accuracy.
A local Hamiltonian is then constructed by projecting $H$ into $\mathcal{P}_I$
\begin{equation}    \label{eq:lno_HI}
    H_{I}
        = \sum_{pq\in \mathcal{P}_I,\sigma}^n
        f^{I}_{pq} a_{p\sigma}^{\dagger} a_{q\sigma}
        + \frac{1}{2} \sum_{pqrs \in \mathcal{P}_I,\sigma\sigma'}^n V_{prqs} 
        a_{p\sigma}^{\dagger} a_{q\sigma'}^{\dagger} a_{s\sigma'} a_{r\sigma'}
\end{equation}
where
\begin{equation}
    f_{pq}^{I}
        = h_{pq} + \sum_{j \notin \mathcal{P}_I}^{N_\mathrm{o}-n_\mathrm{o}} \left(2V_{pqjj} - V_{pjj q}\right),
\end{equation}
which includes a frozen core contribution.
Solving the CCSD amplitude equations %(\ref{eq:ccsd_amp_eqn}) 
with $H_I$ gives the local CCSD amplitudes $T_{Iajb}$ in
$\mathcal{P}_I$ %, where we have transformed one occupied index to $\phi_I$.
and the associated orbital contribution to the correlation energy
\begin{equation}    \label{eq:elnoccsd}
E_I = 
    \sum_{jab \in \mathcal{P}_I} T_{Iajb} (2 V_{Iajb} - V_{jaIb}).
\end{equation}
LNO-CCSD(T) calculations are performed in a similar way, as described
in more detail in Ref.~\onlinecite{Rolik13JCP,Nagy17JCP}.
We note that other fragment-based local correlation methods follow essentially the same idea, but differ only in the definition of the local fragments $I$ and the method for constructing the associated active space of orbitals $\mathcal{P}_I$.

\subsection{LNO-AFQMC}

Adapting the LNO approach for use with AFQMC merely requires an energy expression
analogous to Eq.~(\ref{eq:eccsd_def}). This is straightforward given that
the AFQMC energy is also obtained by left projection onto a trial state, which throughout this work 
we choose to be the HF determinant $\ket{\Phi_0}$. In AFQMC, the ground state
is represented as a statistical average of walkers, each a single Slater
determinant $\ket{\Phi_w}$, with weight $W_w$,
\begin{equation}
|\Psi\rangle = \sum_w W_w |\Phi_w\rangle;
\end{equation}
a review of the AFQMC method with further technical details can be found in Ref.~\onlinecite{marioZhangAbInitioAFQMC}.
The correlation energy is 
\begin{subequations}
\begin{align}
E_\mathrm{c} &= \frac{\sum_{w} W_w E_{\mathrm{c}}^w}{\sum_{w} W_w}, \\
E_{\mathrm{c}}^w &= \frac{\langle \Phi_0 | H-E_{\mathrm{HF}} | \Phi_w\rangle}
    {\langle \Phi_0 | \Phi_w\rangle}.
\end{align}
\end{subequations}
By expanding the walker determinant in the basis of excitations with respect
to the HF trial state,
the correlation energy of a given walker is easily evaluated 
to be
\begin{equation}
\label{eq:eafqmc_def}
E_{\mathrm{c}}^w = \sum_{ijab} G_{ia}^w G_{jb}^w (2V_{iajb}-V_{ibja})
    = \sum_I E_I^w
%C_{iajb} &= G_{ia}G_{jb} \\
%G_{ia} &= \langle \Phi_0 | a_i^\dagger a_a |\Phi_w\rangle
\end{equation}
where
$G_{ia}^w =\langle \Phi_0 | a_i^\dagger a_a |\Phi_w\rangle/\langle \Phi_0 |\Phi_w\rangle$
is the generalized one-particle reduced density matrix.
Eq.~(\ref{eq:eafqmc_def}) clearly has the same form as the CCSD 
expression~(\ref{eq:eccsd_def}). Thus, in LNO-AFQMC, we form localized
occupied orbitals and associated local active spaces $\mathcal{P}_I$
just as in LNO-CCSD; we then perform independent AFQMC calculations in
each local active space $\mathcal{P}_I$ and calculate the contribution
$E_I$ to the correlation energy as an average over walkers,
\begin{subequations}
\begin{align}
E_I &= \frac{\sum_w W_w E_I^{w}}{\sum_w W_w} \\
E_I^w &= \sum_{jab \in \mathcal{P}_I} G_{Ia}^w G_{jb}^w (2 V_{Iajb} - V_{jaIb}).
\end{align}
\end{subequations}
This general form is amenable to almost any flavor of fragment-based local correlation, although here we focus on the LNO framework.
In this work, we perform AFQMC calculations with the phaseless approximation and force bias (hybrid) importance sampling \cite{zhang2003quantum}.

In practice, the missing correlation outside of $\mathcal{P}_I$ can be
included approximately by a composite correction with a lower level of theory, such as MP2. 
To any LNO calculation (CC or AFQMC), we add the correction
\begin{equation}
\label{eq:elnoccsd_mp2correct}
\Delta E^{(2)} = E_\mathrm{c}^{(2)} - E^{(2)}_\mathrm{c,LNO},
\end{equation}
where $E_\mathrm{c}^{(2)}$ and $E^{(2)}_\mathrm{c,LNO}$ are the MP2 correlation energies
in the full orbital space and in the truncated LNO space, respectively.

Summarizing the steps and cost of an LNO calculation, there are three parts.
\begin{enumerate}
    \item Full-system MP2, which is required by both the LNO construction [Eqs.~(\ref{eq:DIij})--(\ref{eq:lno_occ})] 
and the MP2 composite correction [Eq.~(\ref{eq:elnoccsd_mp2correct})] and scales as $O(N^5)$.
    \item $N_{\mathrm{o}}$ independent integral transformations, which are required by the local Hamiltonian construction [Eq.~(\ref{eq:lno_HI})] and scale as $O(N^4 n)$ each, but embarrassingly parallel in $N_{\mathrm{o}}$.
    \item Independent correlated calculations of all local Hamiltonians, which scale as
    $N_\mathrm{o}$ times the cost of a calculation at the desired level of theory in the
local active space, i.e., $n^4$ for AFQMC, $n^6$ for CCSD, and $n^7$ for CCSD(T). 
\end{enumerate}
For moderately sized systems, the high-level correlated calculation in step (3)
dominates the computational cost, which leads to an overall cost that scales
linearly with the system size.  This will be the case for all the systems we
use to benchmark our method in this work.
As the system size increases, the first two steps whose cost scales superlinearly with the system size $N$ eventually become the computational bottleneck.
Although not explored in this work, many numerical techniques such as local
domain-based approximations~\cite{Rolik13JCP} and Laplace transform
methods~\cite{Nagy17JCP} have been exploited to make these steps linear scaling as well.
Such advances can be straightforwardly used with the LNO-AFQMC approach described here.

\subsection{Computational scaling of AFQMC and LNO-AFQMC}

In AFQMC, energies are obtained by averaging over a trajectory that samples the wavefunction.
With force bias (hybrid) importance sampling, the cost of each propagation step scales as $N^3$.
Local energy evaluation scales as $N^4$ but is performed less frequently; for moderately sized systems,
including those studied here, the total cost is dominated by propagation and thus scales effectively as $N^3$.
However, for the following scaling analysis, we will assume the worst-case scenario where local energy evaluation dominates,
although we note that its scaling can be reduced to $N^3$ using
integral compression~\cite{Morales2019afqmc,Chan2019afqmc,Reichman2020afqmc} or localized orbitals~\cite{Friesner2022afqmc}.

The above scalings are for a trajectory with a fixed number of iterations $N_t$, but how does $N_t$
scale with system size $N$? Assuming the variance of the total energy is proportional to system size, the stochastic error after $N_t$ iterations is $\sqrt{\sigma^2/N_t} \propto \sqrt{N/N_t}$.
Therefore, to achieve a fixed absolute error requires $N_t\propto N$, but to achieve a fixed relative error
(i.e., error per electron) requires $N_t \propto 1/N$.
Thus, for a calculation dominated by $N^4$-scaling energy evaluation, the final cost of AFQMC scales as $N^5$ for fixed absolute error
and as $N^3$ for fixed relative error~\cite{Foulkes01}.

In LNO-AFQMC, the computational scaling of a single propagation step (including local energy evaluation) is effectively reduced from $N^4$ to
$N n^4$, where $n$ is independent of system size because each fragment is treated independently.
This reduction by a factor of $N^3$ implies that the cost of LNO-AFQMC scales as $N^2$ for fixed absolute error and is independent of $N$ for
fixed relative error (to be compared to $N^5$ and $N^3$, respectively, for the canonical algorithm).

Separately, one needs to consider the biases inherent in AFQMC results due to Trotter error and truncation of Cholesky decomposition of the two-electron integrals. 
In both cases we expect the errors to increase with the size of the system for a fixed time step and Cholesky threshold. 
Thus, to obtain a constant error with increasing system size, one has to run calculations with smaller time steps and Cholesky threshold. 
In LNO-AFQMC these shortcomings are mitigated because each individual calculation contains a small number of electrons. 
Empirically, we find that in LNO-AFQMC calculations, we can use a Cholesky threshold that is an order of magnitude larger than in canonical AFQMC calculations without seeing a noticeable error in the final results (see Section~\ref{sec:results}). 
Similarly, we find that in large AFQMC calculations, we need to use smaller Trotter time steps to avoid large biases.

\section{Results}\label{sec:results}
%In this section, we show results of test calculations done to evaluate the performance of LNO-AFQMC. 
In this section, we present the results of LNO-AFQMC calculations for total energies and isomerization reaction energies. 
Tight convergence of total energies is naturally more difficult to achieve than that of energy differences.
%and their convergence with the adjustable thresholds $\epsilon_o$ and $\epsilon_v$. 
All HF, MP2, and CC calculations were performed using PySCF~\cite{PySCF}, and all AFQMC calculations were performed using Dice~\cite{Dice} with a Trotter timestep of 0.005 a.u., unless specified otherwise. 
The geometries used for the total energy calculations are provided in the Supplemental Material. For all calculations, Dunning correlation-consistent basis sets [cc-pVXZ (where X=D,T,Q) or aug-cc-pVDZ]~\cite{Dunning1,Dunning2,Dunning3} were used. The core electrons were kept frozen in all the calculations. 

Occupied orbitals were localized using the Pipek-Mezey method~\cite{PipekMezey}. 
Following previous work~\cite{Nagy18JCTC}, a ratio of $\epsilon_\textrm{o}/\epsilon_\textrm{v}=10$ was fixed to reduce the number of variables, and we generally test
the range from $\epsilon_\textrm{v}=10^{-4}$ (loosest) to $\epsilon_\textrm{v}=10^{-6}$ (tightest).
In the LNO-AFQMC fragments, Cholesky decomposition was performed with a threshold error of $10^{-4}$, while for the reference AFQMC calculations, a more stringent threshold error of $10^{-5}$ was employed. %\tcb{Confused by this sentence}. 
%Even though there can be multiple systematic biases associated with AFQMC, we estimate these to be less than the stochastic error \tcb{Even with a single determinant trial?}. 
All LNO-AFQMC calculations were converged to $1$~mHa stochastic error, which requires converging the correlation energy contribution from each fragment to a stochastic error that is smaller by a factor of $\sqrt{N_\mathrm{o}}$, where $N_\mathrm{o}$ is the number of fragments (equal to the number of occupied orbitals). 
For canonical AFQMC calculations, a stochastic error of 1 mHa can be achieved for small molecules, but not for large basis sets and large molecules without significant computer resources.

\subsection{Total energies}

\begin{figure}
    \centering
    \includegraphics[width=\columnwidth]{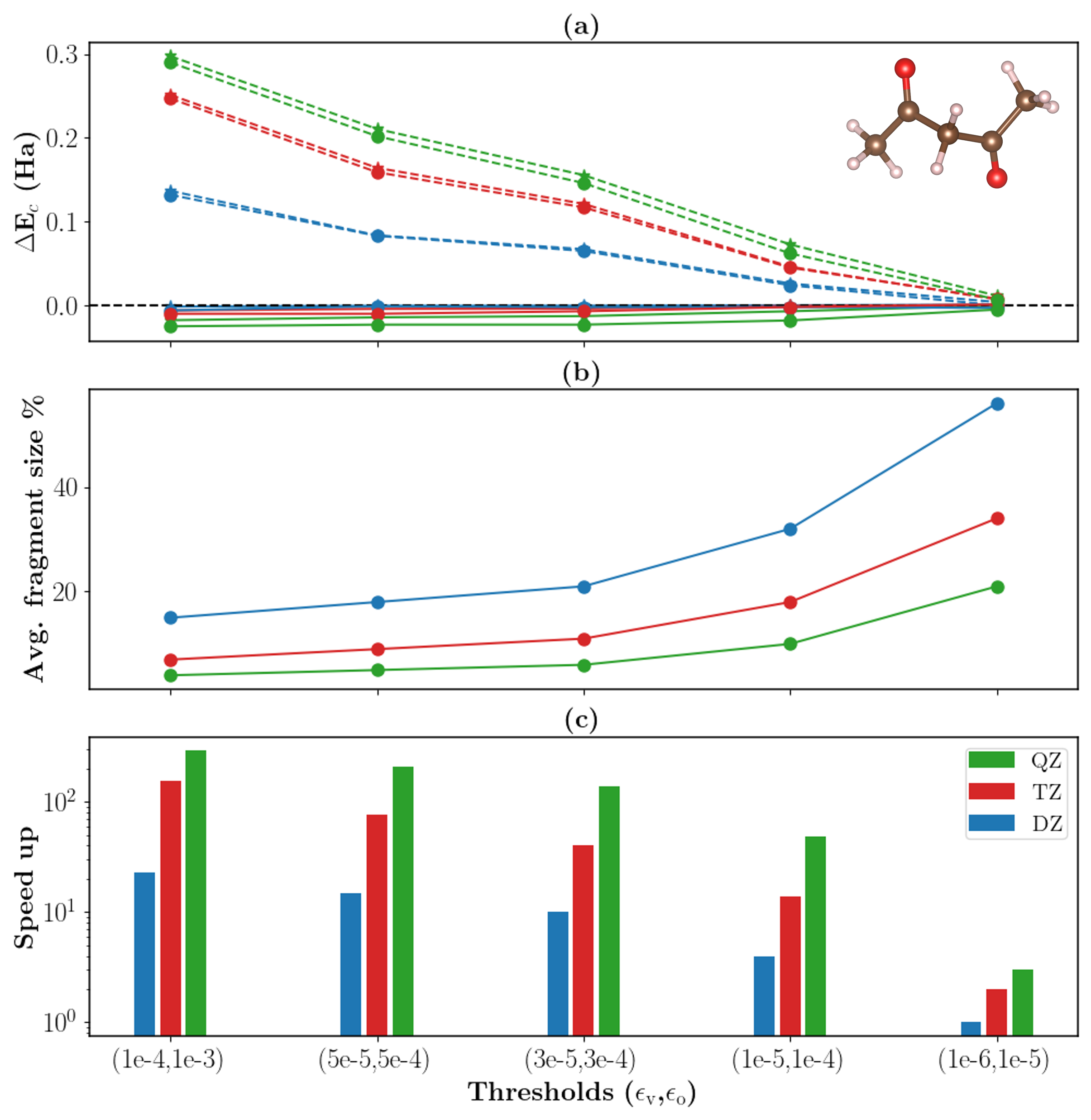}
    \caption{Performance of LNO-AFQMC and LNO-CCSD(T) for the total correlation energy of the acectylacetone molecule shown in the inset. (a) Convergence of the correlation energy of LNO-AFQMC (circles) and LNO-CCSD(T) (star), with (solid lines) and without (dashed lines) MP2 corrections in the DZ (blue), TZ (red), and QZ (green) basis sets. (b) The average number of orbitals in each LNO fragment as a percentage of the total number of molecular orbitals. (c) Speed-up of LNO-AFQMC compared to canonical AFQMC, where the timing of LNO-AFQMC is reported as the sum of times for all fragments.}
    \label{fig:LNOCCSD(T)}
\end{figure}

Total energy calculations were performed on acetylacetone, which is small enough (40 valence electrons) to allow canonical CCSD(T) and AFQMC calculations in the DZ (131 orbitals), TZ (315 orbitals), and QZ (618 orbitals) basis sets for benchmarking purposes.
In Fig.~\ref{fig:LNOCCSD(T)}(a), we show the convergence of the correlation energy with LNO threshold using different basis sets.
Without the MP2 correction, the correlation energy error is large but progressively converges with tighter thresholds. 
The convergence is significantly accelerated with the MP2 correction.
Specifically, in the DZ basis (with the MP2 correction), the correlation energy error is 6 mHa with loose thresholds and 3 mHa with tight thresholds, the latter of which is comparable to the stochastic error of calculation. 
Similarly, in the TZ  and QZ basis sets, the error decreases from 10 mHa to 1 mHa and from 25 mHa to 3 mHa with increasingly tight thresholds.
We note that all of these errors amount to less than 2\% of the total correlation energy.
This convergence behavior is almost identical to that from LNO-CCSD(T), results of which are also shown in Fig.~\ref{fig:LNOCCSD(T)}(a), confirming the straightforward transferability of the LNO methodology.

As discussed above, the key advantage of the LNO methodology is the reduction in the number of orbitals that need to be correlated within each fragment.
In Fig.~\ref{fig:LNOCCSD(T)}(b), we show the average size (total number of orbitals) of the LNO fragments of acetylacetone with the DZ, TZ, and QZ basis sets. 
Even with the tightest threshold, the average number of fragment orbitals is 56\%, 34\%, and 21\% of the total number of orbitals in DZ, TZ, and QZ basis sets, respectively, showing that the method is particularly advantageous for larger basis sets.
For correlated methods with polynomial scaling, these reductions in the number of orbitals lead to huge savings in the computational cost.
This behavior is shown in Fig.~\ref{fig:LNOCCSD(T)}(c), where we report the speed up, calculated as the ratio of time taken for the canonical AFQMC calculation and the total LNO-AFQMC calculation. The time taken for LNO-AFQMC is reported as the sum of times for all fragment calculations, but because these calculations are independent, the walltime can be reduced by a factor approximately equal to the number of fragments if these calculations are performed in parallel.

In the DZ basis set with a target stochastic error of 1~mHa, the LNO method with the loosest threshold accelerates the calculations by a factor of about 23; with the tightest threshold, the time becomes comparable to canonical AFQMC. 
Results are even more encouraging in larger basis sets, where the number of orbitals per fragment is a smaller fraction of the total and the speed-up is therefore more significant. In the QZ basis, the speed-up ranges from almost 300 to 3 with an increasingly tight threshold, and even in the latter case, the stochastic error of LNO-AFQMC was converged to 1~mHa while that for canonical AFQMC could only be converged to 2~mHa.
Finally, we note that these timings and speed-ups pertain to a relatively small molecule for which canonical AFQMC calculations are feasible. As shown in the Supplemental Material, the speed-up becomes even more pronounced for larger molecules [melatonin (90 valence electrons) and penicillin (128 valence electrons)] where obtaining canonical benchmark results in large basis sets becomes impractical. We conclude that, while specific timings are influenced by acceptable errors, molecular size, basis set, and hardware, LNO-AFQMC demonstrates better efficiency compared to its canonical counterpart and scales effectively to larger systems with larger basis sets.

\subsection{Isomerization energies}\label{subsec:isomerization}

\begin{figure}
   \centering
  \includegraphics[width=\columnwidth]{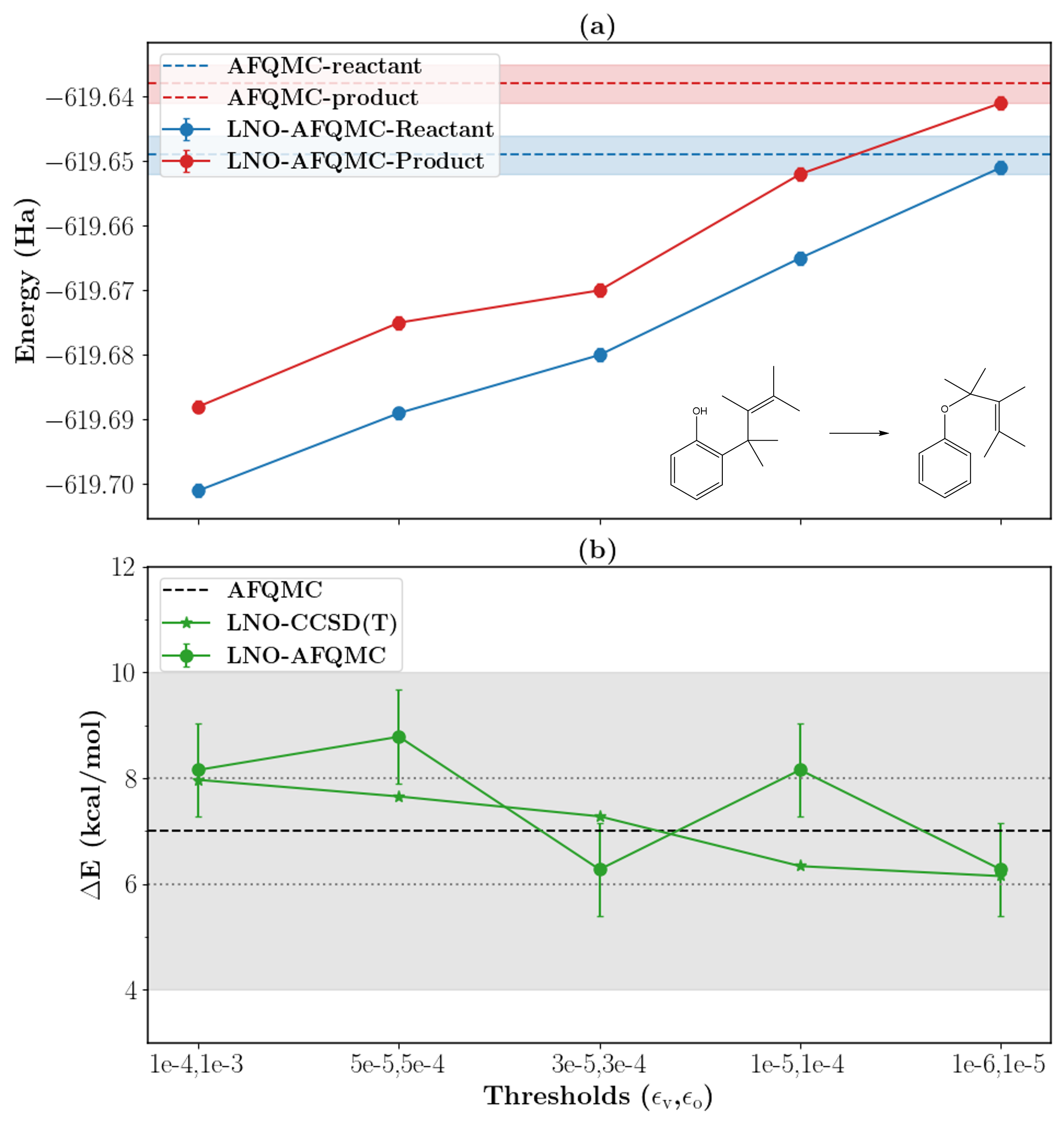}
  \caption{ 
    Performance of LNO-AFQMC (with MP2 corrections) for the isomerization energy of reaction 10 from the ISOL database [shown in (a)] with the cc-pVTZ basis set. Convergence of the total energy of the reactant and product (a) and their energy difference (b) as a function of threshold. The red, blue, and grey shaded regions indicate the stochastic error of the canonical AFQMC calculations, and the dotted lines indicate $\pm 1$~kcal/mol around the average canonical AFQMC isomerization energy. 
    }
  \label{fig:rn10_tz}    
\end{figure}

\begin{table}[]
\begin{ruledtabular}
\caption{Isomerization energy (kcal/mol) of reactions 3, 9 and 10 (see SM for figures of reactions) obtained using LNO-AFQMC, AFQMC, and CCSD(T) in the cc-pVTZ basis set.
For reaction 10, because of its large size, we give the result from LNO-CCSD(T) with the tightest threshold.}
\label{tab:rn_tz}
\begin{tabular}{cccc}
    & Reaction 3      & Reaction 9       & Reaction 10     \\
\hline
LNO-AFQMC (loosest) & 8.8(9) & 22.6(9) & 8.2(9) \\
LNO-AFQMC (tightest) & 8.8(9)  &  22.0(9) &  6.3(9) \\
AFQMC     & 9.4(9) & 20(1)   & 7(3)   \\
CCSD(T)   & 8.77   & 21.63   & 6.16 
\end{tabular}
\end{ruledtabular}
\end{table}

The relative energy differences between structures during chemical reactions are often more significant than their absolute energies. 
In order to assess the performance of LNO-AFQMC for chemical reactions, we studied three isomerization reactions from the ISOL database, which contains 24 isomerization reactions of large organic molecules~\cite{ISOL-old, ISOL24/11}.
As a case study, we focus on reaction 10, shown in the inset of Fig.~\ref{fig:rn10_tz}(a). 
The molecule has 15 heavy atoms, 82 electrons, and 715 orbitals in the cc-pVTZ basis set; it has one of the smallest isomerization energies in the ISOL database, and it is the most challenging of the three reactions we study.
As shown in Fig.~\ref{fig:rn10_tz}(a), with increasingly tight LNO thresholds, the error in the total energy (with the MP2 correction), for both the reactant and product, decreases from about 50~mHa to 2--4~mHa.  
Importantly, the error is very similar for both the reactant and product at a given threshold, such that the energy difference (the isomerization energy) is almost independent of the threshold.
In fact, the isomerization energy is always within the relatively large stochastic error bars of the canonical AFQMC calculation, which predicts an isomerization energy of $7 \pm 3$~kcal/mol.
From our tightest threshold, the LNO-AFQMC isomerization energy is predicted to be $6.3\pm 0.9$~kcal/mol; for comparison, LNO-CCSD(T) isomerization energy in the same basis set was calculated as 6.16~kcal/mol.

We have performed the same calculations for reactions 3 and 9, and detailed results for all reactions are presented in the Supplemental Material.
In all cases, we find that energy differences converge significantly faster than total energies and are always within the stochastic error of the canonical AFQMC result, even for loose thresholds. In Tab.~\ref{tab:rn_tz}, we report LNO-AFQMC isomerization energies for all three reactions obtained using the loosest and tightest threshold, compared to canonical AFQMC and CCSD(T) in the cc-pVTZ basis set. 
The LNO-AFQMC results are in good agreement with those from canonical AFQMC and required only a fraction of the cost. Employing the loosest threshold had a speedup of 158, 233 and 43 for reactions 3, 9 and 10, respectively (and recall that, for reaction 10, the canonical AFQMC calculation had only reached convergence within a stochastic error of 3~kcal/mol). Even with the tightest threshold, LNO-AFQMC was still faster than the canonical counterpart by a factor of 3 for reaction 3 and 9 and comparable for reaction 10 despite converging to a smaller stochastic error.
Moreover, the agreement with (LNO-)CCSD(T) is quite good, reflecting the single-reference character of these organic molecules.

Interestingly, when using a Trotter timestep of 0.005 a.u., the canonical AFQMC energy calculation for the reactant of reaction 9 showed a notable difference of 7~mHa compared to the LNO-AFQMC calculation. When the timestep was halved, this difference reduced to 4~mHa. This behavior can be attributed to the scaling of Trotter error with system size. However, the LNO-AFQMC approach effectively circumvents these biases by employing fragment calculations that are considerably smaller in scale compared to the overall system. 
Similar sensitivity to Trotter timestep was observed for the product of reaction 3 in the aug-cc-pVDZ basis set. Notably, the AFQMC calculation with a timestep of 0.0025 a.u.\ closely aligns with the tightest LNO-AFQMC result obtained using a step size of 0.005 a.u., which demonstrates the reliability of the local method.

\section{Conclusion}
\label{sec:conc}
In this work, we have introduced local correlation into the AFQMC framework, specifically via the use of local natural orbitals.
The LNO framework provides an efficient truncation of the basis set, which makes it possible to perform LNO-AFQMC calculations with larger basis sets than is possible with canonical AFQMC.
Notably, energy differences converge much more rapidly than total energies, which makes this method especially promising for applications in chemistry. 
In the future, LNO-AFQMC will be extended to study open-shell and strongly correlated, multi-reference systems. 
To achieve this, we will adapt LNO-AFQMC for use with multi-determinantal trial states~\cite{mahajan2021,Sharma2022afqmc},
for which there is no unique set of occupied orbitals and Eq.~(\ref{eq:eafqmc_def}) no longer holds.

\section*{Acknowledgements}
This work was supported by the National Science Foundation under Grant Nos.~CHE-2145209 (J.S.K.),
OAC-1931321, and CHE-1848369 (H.-Z.Y. and T.C.B.), 
and by a grant from the Camille and Henry Dreyfus Foundation (S.S.). This work utilized resources from the University of Colorado
Boulder Research Computing Group, which was supported by
the National Science Foundation (Award Nos. ACI-1532235 and
ACI-1532236), the University of Colorado Boulder, and Colorado
State University. 

\bibliography{references}

\end{document}